\documentclass[12pt]{iopart}
\usepackage{graphicx}

\newcommand {\reals} {{\rm I\!R}}

\newcommand {\bU} {\mbox{\boldmath $U$}}
\newcommand {\bV} {\mbox{\boldmath $V$}}

\newcommand {\bX} {\mbox{\boldmath $X$}}

\newcommand{\calA}{{\cal A}}

\newcommand{\calJ}{{\cal J}}

\begin{document}

\title[Bose--Einstein Condensation in Large Deviations ]{Bose--Einstein
Condensation in the Large Deviations 
Regime with Applications to Information System Models}

\author{Neri Merhav$^{(1)}$ and Yariv Kafri$^{(2)}$}

\address{(1) Department of Electrical Engineering, Technion, Haifa 32000, Israel. 
(2) Department of Physics, Technion, Haifa 32000, Israel}

\begin{abstract}
We study the large deviations behavior 
of systems that admit a certain form of a 
product distribution, which is frequently encountered
both in Physics and in various information system models. First, to fix ideas, we
demonstrate a simple
calculation of the large deviations 
rate function for a single constraint (event).
Under certain conditions, the behavior of this function is shown to exhibit an 
analogue of Bose--Einstein condensation (BEC). More interestingly, 
we also study the large deviations rate function associated with two
constraints (and the extension to any number of constraints is conceptually
straightforward).
The phase diagram of this rate function is shown to exhibit as many as 
seven phases, and it suggests a two--dimensional generalization of the notion
of BEC (or more generally, a multi--dimensional BEC).
While the results are illustrated 
for a simple model, the underlying principles are actually rather general. 
We also discuss several applications and implications pertaining to information system models.
\end{abstract}

\maketitle

\section{Introduction}

While the theory of statistical physics is traditionally concerned with typical or almost 
typical events, the closely related theory
of large deviations deals with rare events whose probabilities are
exponentially small in the size of the system.
More precisely, large deviations theory is concerned with the exponential
decay rate of probabilities of certain rare events, 
as the number of observations grows without bound.
In statistical mechanics, there has always been some interest 
in the statistics of rare events (the Kramers escape problem being one example \cite{vanKampen}). 
More recently, the interest in rare events has grown due to several applications. 
For example, in many cases it is important to know the probability of an
extinction event in non--equilibrium models of
epidemics \cite{Mer1,Mer2}. Another example is the measurement of fluctuation theorems 
\cite{jar1,jar2}, such as the Jarzynski equality, which rely on probing rare events. 
Interest in rare events has also emerged recently in 
the statistics of records and other stochastic processes \cite{Krug,Satya,Satya2}. 
Finally, the calculation of large deviations is a natural framework 
within which one can define non--equilibrium free energy analogues \cite{Der1,Der2,Der3,Der4}.   

In this paper, we consider large deviations pertaining to product measures. 
In particular, we focus on the probability that some quantity of relevance
would exceed a certain threshold. 
The paper contains two parts. In the first, we give a discussion with a
tutorial flavor, which focuses on the calculation of the probability of a 
simple single large deviations events. Our aim, in this part, is to point out,
for a non--expert reader, two important aspects: The first is the relation between 
large deviations theory and conventional statistical physics, and the second
is the fact that phase transitions can be observed 
in the large deviations regime even in simple systems with no interactions,
where phase transitions are not expected in the usual regime, of analyzing the
typical behavior of the system. In particular, for product 
distributions of a certain form, a direct analogue of Bose--Einstein condensation 
(BEC) can be observed. Following this tutorial part, we turn to the second part of 
the paper, where we present results that extend the calculations of a single event 
to accommodate two simultaneous events,
and the further extension to any finite and fixed number of events is then conceptually
obvious. We show that even 
when the two events are physically closely related, the phase diagram can exhibit 
as many as seven different phases. This means that the large deviations
point--of--view actually
suggests a multi--dimensional extension of the notion of BEC. 
Furthermore, we compare phase diagrams of large deviations rate functions pertaining to inequality events
to those of equality events and it turns out that these two phase diagrams are very different.

To fix ideas, we first illustrate the results for a 
simple hopping model (closely related to the models studied in
\cite{Mukamel,Pulk,Ange}), but they remain valid for fairly general forms of product
distributions, and as such, apply to many physical systems and information
system models. Examples of these range from black--body 
radiation (for a related calculation of large deviations in ideal quantum
gases, see \cite{LebSph}), zero-range processes 
(in and out of equilibrium) \cite{Evans00}, Jackson
networks, which emerge in queuing theory
(and which are essentially analogous to 
zero range processes, but with no conservation of the particle number) 
\cite{Jackson57,Jackson63}, driven--diffusive systems \cite{Evans96}, and many others. 
These product distributions also arise in additional engineering applications. 
For example, this is the natural distribution for a one--way Markov chain, 
which is defined by an ordered set of states, where the 
only allowed transitions from each state are the self--transition
and a transition to the next state.
One--way Markov processes are commonly used in statistical modeling for a 
wide spectrum of application areas, including information theory, 
communications and signal processing (see Section 5 for details). 

The outline of the remaining part of this paper is as follows: In Section 2, we illustrate our 
results, without the detailed derivation, using a simple one--dimensional hopping model, 
which may describe transport in a disordered medium. 
In Section 3, we derive general results for the large deviations rate
function of a single constraint. In Section 4, we extend the derivation
to incorporate two constraints, and then display the corresponding phase diagram. 
finally, in Section 5, we discuss several applications to information system models.

\section{Informal Illustration of the Results}

\begin{figure}[ptb]
\begin{center}
\includegraphics[width=8cm ]{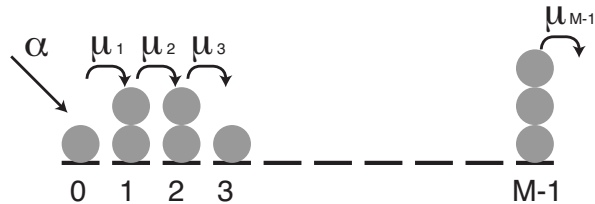}
\end{center}
\caption{An illustration of the hopping model. Particles enter into the system
from the left at rate $\alpha$. If site $i$ is occupied, a particle is
transferred to site $i+1$ at rate $\mu_i$. Particles leave the system at rate
$\mu_{M-1}$ on the right--hand side. For a closely related model see, for example, \cite{Mukamel}.}
\label{model}%
\end{figure}

Throughout this paper, we consider systems whose steady-state behavior admits
a probability distribution of the product form
\begin{equation}
\mbox{P}(\{ n_i \})=\frac{1}{Z} \prod_{i} p_i^{n_i},
\label{proddist}
\end{equation}
where $n_i$ is the number of ``particles'' in ``lattice site'' $i$ of the
system and
$Z$ is a normalization constant. This means that $
\{n_i\}$ are independent geometric random variables with parameters $
\{p_i\}$.
The immediate relevance of this model is the distribution of the occupation numbers $\{n_i
\}$
of the various energy levels in the grand canonical ensemble of an ideal boson gas,
where $p_i=ze^{-\beta\epsilon_i}$, $z$ being the fugacity, $\beta$ -- the inverse temperature,
and $\{
\epsilon_i\}$ are the corresponding energy levels. Other natural applications of this model, which were
mentioned briefly in the Introduction, will be reviewed in detail in Section 5.
One can easily generalize our results 
to the case where each factor in the product of (\ref{proddist}) is $p_i^{n_i}/n_i^b$.
For the sake of simplicity, however, we confine 
ourselves throughout to the form (\ref{proddist}), for which $b=0$.
For concreteness and intuition, we focus on a particularly simple 
dynamical model with such a steady--state distribution 
(for related models, see \cite{Mukamel,Pulk,Ange}). The model is defined on a
one--dimensional lattice, with $M$ sites, labeled by $i=0,1, \ldots, M-1$. 
A configuration of the system is defined by the number of particles 
$n_i=0,1,\ldots,M-1$ at each site. The evolution is governed by random sequential 
dynamics defined by the following rules: Particles enter into the system via site $i=0$ at 
rate $\alpha$ (there is no exclusion between the particles). If at site $i$
$n_i>0$,
a particle is transferred from site $i$ to site $i+1$ at rate $\mu_i$. 
At site $i= M-1$, particles leave the lattice at rate $\mu_{M-1}$. 
The model, as illustrated in Fig. \ref{model}, is therefore non--conserving 
only at the edges of the system. It can be considered as a simple model for transport in a 
disordered medium or, more pictorially, as a model of
customers being served along a sequence of $M$ consecutive queues, from left to right.
In the latter case, each site represents a server. In the realm of queuing
network theory, this model is a 
specific example of a Jackson network \cite{Jackson57,Jackson63},
and in steady--state, it admits a product of distributions of geometric random variables, provided that the 
rate at which particles flow into the system is small enough, so that the
system does not overflow, namely, in this case, $\alpha < \min_i \{ \mu_i \}$. 
Specifically, the steady--state probability of a configuration $(n_0,\ldots,n_{M-1})$, in this example, is given by
\begin{equation}
\mbox{P}(n_0,n_1,\ldots,n_{M-1})=\frac{1}{Z} \prod_{i=0}^{M-1} \left(\frac{\alpha}{\mu_i}\right)^{n_i},
\label{proddist2}
\end{equation}
which falls in the framework of (\ref{proddist}) with $p_i=\alpha/\mu_i$. 

Our interest is in calculating the probability of a certain large--deviations
(rare) event ${\bf
X}$. A simple example of such an event,
customarily considered in large deviations theory, 
is that the total number of particles in the lattice exceeds some threshold,
that is, ${\bf X}=\{(n_0,n_1,\ldots,n_{M-1}):~~\sum_{i=0}^{M-1} n_i \geq N\}$. 
Consider the thermodynamic limit where both $N$ and $M$ grow without
bound, such that their ratio $N/M\equiv U$ is kept fixed. If $U$ exceeds
a minimum value, given by its average value, and which we shall denote $U_{\min}$, this event becomes
asymptotically rare, and its
probability, $\mbox{Pr}\{{\bf X}\}$ decays with $M$ asymptotically exponentially at the same
as $\exp[-M\cdot J(U)]$,
where $J(U)$ is referred to as the {\it large deviations rate function} in
large deviations theory.
Our interest will therefore be primarily in 
the evaluation of $J(U)$.
Another relevant question would be about
characterizing those
configurations of the system that dominate $J(U)$. In other words, 
given that the event ${\bf X}$ has occurred, what are the system
configurations that one is likely to observe?

In general, $J(U)$ may not be a smooth
function. It may exhibit singularities (e.g.,
discontinuities in the derivatives of $J(\cdot)$) at some value $U=U_c$
(perhaps even at more than one such value).
In the sequel,
these will be referred to as {\it phase transitions} 
of the large deviations
rate function. These phase transitions may be manifested not merely in
possible singularities of the function $J$, but more interestingly, in
condensation phenomena pertaining to the dominant configurations of
the large deviations event in question. For example,
under certain conditions on the asymptotic behavior of rates $\{\mu_i\}$
(analogous to conditions on the density of states in BEC), for $U>U_c$, the 
dominant configurations become condensed: Although the total number of sites
$M$ grows without bound, a macroscopic fraction of
the particles reside only in one of them. Loosely speaking,
the particles are essentially jammed 
at the site (or server) with the slowest exit rate \cite{Evans96}. 
The value $U_c$ is analogous to the critical density in the ordinary 
BEC transition. For large deviations events of the type $\sum_i n_i \geq N$, the rate function exhibits 
an additional phase transition: When $U < U_{\min}$, the event in question 
is no longer rare and so $J(U)=0$. This is a direct result of looking at an
event defined by an inequality constraint rather than an equality constraint. 
If one considers instead a constraint of the form $\sum_i n_i = N$, one would
find two phases only, a condensed phase ($U > U_c$) and a non--condensed phase
($U < U_c$), and not three. In the sequel, we will elaborate more
on this difference between equality constraints and inequality constraints.

Interestingly, condensation phenomena occur also for other 
constraints defined in terms of various 
linear combinations of $\{n_i\}$. For example, $\hat{T}{\bf \equiv} \sum_i n_i/\mu_i$ is a plausible
estimate for the total time that a particle would spend in the system, because
$n_i/\mu_i$ is the expected time that each particle spends at site $i$ before
being moved, in its turn, to the right. 
Consider now the event 
$\hat{T} \geq M\cdot V$. The large deviations behavior of this event also exhibits two phase
transitions, one at $V= V_{\min}$, where $J(V)$ ceases to be
identically zero and becomes strictly positive, and the other at
$V=V_c$, from the non--condensed to the condensed phase,
with $V_c$ depending on the rates in the system. 
Once again, in the condensed phase, particles essentially jam at the site with
the smallest exit rate.

More surprising and interesting is the phase diagram obtained for the joint
probability of two rare events pertaining to 
two different linear combinations of $\{n_i\}$, say, 
$\mbox{Pr}\{\sum_i n_i \geq M\cdot U,~\sum_in_i/\mu_i \geq M\cdot V\}$, 
which decays exponentially according to $\exp\{-M\cdot J(U,V)\}$ for some
rate function $J(U,V)$.
As we show in the sequel, even if the two constraints are physically closely
related, the phase diagram of the large deviations rate function $J(U,V)$
has a very rich phase diagram with as many as seven different phases. 
We find three distinct types condensed phases: one for each one of the 
individual events and a third one for their combination, which gives rise to
the notion of a {\it two--dimensional condensation}. Furthermore, the phase diagram 
associated with the corresponding equality constraints, $\sum_i n_i = MU$ and
$\sum_i n_i/\mu_i = MV$, is dramatically different from 
that of the inequality constraints, with two phases only rather than seven.  Note, that this two dimensional condensation is very different in nature from that considered in the context of two, distinct, conserved quantities \cite{EvansHanney1,EvansHanney2,Gross}.
The two--constraint problem, in its general form, is the focus of the main part of the paper. 
In the next section, we present a detailed derivation of the results for the single constraint problem.

\section{A Single Constraint}

As mentioned in the Introduction, we begin with 
a simple single constraint, 
assuming that one has a product distribution of the form of eq.\
(\ref{proddist}). 
Referring to the terminology of particles and sites, from the example of
Section 2, 
consider first the probability of the event 
that the number of particles in the system is larger than some threshold,
$\sum_i n_i \geq M\cdot U$. The large deviations evaluation of this probability 
is typically done using the Chernoff bound. Specifically,
consider the following chain of inequalities:
\begin{eqnarray}
\mbox{Pr}\left\{\sum_i n_i \geq MU\right\}&\le&
\left<z^{\sum_i n_i-MU}\right>~~~~~~~~z\ge 1\nonumber\\
&=&z^{-MU}\prod_i \frac{1-p_i}{1-z p_i}\nonumber\\
&=&\exp\left\{-M\left[U\ln z -
\frac{1}{M}\sum_i\ln\left(\frac{1-p_i}
{1-z p_i}\right)\right]\right\},
\label{Chern1}
\end{eqnarray}
where the angular brackets denote an expectation with respect to the distribution
of eq.\ (\ref{proddist}).
The tightest bound, which gives the large-deviations rate function, is
obtained by minimization of the Chernoff bound
over $z$, or equivalently, by maximization of the bracketed expression at the
exponent:
\begin{equation}
\label{Jsimple}
J(U)=\sup_{z\ge 1}\left[U\ln
z-\lim_{M\to\infty}\frac{1}{M}\sum_{i=0}^{M-1}\ln\left(\frac{1-p_i}{1-zp_i}\right)\right],
\end{equation}
provided that the limit exists for all $z\ge 1$.

Note that the {\it Chernoff parameter} $z$, that undergoes optimization, is 
almost equivalent to the fugacity $z$ 
of the grand--canonical ensemble, which controls the expected number of particles 
in the system, and the minimization of the bound is parallel to the 
usual saddle--point evaluation pertaining to the grand partition function.
The only difference comes about since the Chernoff bound is 
concerned with the probability of the inequality event 
$\sum_i n_i \geq MU$, as opposed to the event $\sum_i n_i = MU$, which
defines the canonical ensemble with $N=MU$.
This implies that one is interested in $z \geq 1$ 
and when the number of particles is below its average value, $J=0$. 
For rare events (with $J>0$), the distinction between 
$\mbox{Pr}\left\{\sum_i n_i \geq N\right\}$ and $\mbox{Pr}\left\{\sum_i n_i = N\right\}$ 
becomes meaningless in the limit of large $N$ 
due to the exponential decay of the probability with $M$. 
With this analogy, clearly, in the limit of large $M$ (or equivalently $N$), 
the bound gives an asymptotically exact value of the rate function $J$ (see,
e.g., \cite{DZ93}). 
In other words, the calculation of the large deviations 
probability of a rare event is essentially identical to a change of ensembles 
in traditional statistical physics, with the rate function 
$J$ playing the role of a free energy. An extra, somewhat trivial, 
phase occurs due to the constraint taking the form of an inequality 
and not an equality of the form $\sum_i n_i = N$. In the latter case, 
the phase with $J=0$ would not exist. 
With this in mind, what follows in the next paragraph is standard.

As mentioned earlier, in order to proceed from eq.\ (\ref{Jsimple}), we must 
assume that the limit in eq.\ (\ref{Jsimple}) exists. We will assume that there exists a density function
$g(t)\ge 0$, integrating to unity, such that in the limit of $M\to\infty$, the fraction of $\{p_i\}$
that fall between $t$ and $t+\mbox{d}t$, tends to $g(t)\mbox{d}t$ for all $t\in(0,1)$.
Performing a saddle--point approximation on eq.\ (\ref{Chern1}) gives the following equation
for the optimum choice of $z$:
\begin{equation}
U= \lim_{M\to\infty}\frac{1}{M}\sum_{i=0}^{M-1}\frac{zp_i}{1-zp_i} = \int_0^{p_m}
\frac{z t g(t)\mbox{d}t}{1- z t},
\end{equation}
where $p_m=\max_i p_i$,
and where it is assumed that $p_m$ is attained by the 
same $i$ (say, $i=0$ without loss of generality) for all $M$.\footnote{
Note that for $p_i=e^{-\beta\epsilon_i}$, this is exactly the classical equation that underlies the BEC.}
Let us denote
\begin{equation}
\label{eqvarrho}
\bU(z)\equiv z\int_0^{p_m}\frac{tg(t)\mbox{d}t}{1-z t}.
\end{equation}
We therefore have to solve the equation $U=\bU(z)$,
where the solution $z$ is sought in the range $[1,1/p_m)$. Now, in analogy
with BEC, if $g(p_m)=0$
and $\lim_{t\uparrow p_m} g(t)/(p_m-t)^\chi$ is positive and finite
for some $\chi > 0$, then $\bU(1/p_m) < \infty$, and 
so, the large deviations behavior exhibits a condensation. In other words,
as long as $U$ is below the the critical density:
\begin{equation}
U_c=\bU(1/p_m)\equiv
\int_0^{p_m}\frac{tg(t)\mbox{d}t}{p_m-t},
\end{equation}
there is no condensation, while for $U > U_c$,
condensation takes place.
This means that the large deviations event in question is dominated by
realizations for which $n_0/N$ is about $U-U_c > 0$, while
all other states have negligible relative contributions. Here $n_0$ is the occupation at the site $i=0$,
corresponding to $p_m$. Denoting by $U_{\min}=\bU(1)$, 
the average particle density in the system, the corresponding large deviations rate function is given by:
$$J(U)= \left\{\begin{array}{ll}
0 & U < U_{\min} \\
U\ln z-\int_0^{p_m}\mbox{d}tg(t)
\ln\left(\frac{1- t}{1-z t}\right) & U_{\min}\leq U < U_c\\
U\ln\left(\frac{1}{p_m}\right)-\int_0^{p_m}\mbox{d}tg(t)
\ln\left(\frac{1- t}{1-t/p_m}\right) & U \ge U_c
\end{array}\right. $$
so that $\mbox{Pr}\{\sum_i n_i \geq N\}$ is of the exponential order of
$\exp\{-MJ(U)\}$
and the large deviations rate function exhibits 
three phases: the first is where $U$ is below the average value, the second is the
non-condensed phase, and the third is the condensed phase. 

The above derivation can be extended quite
straightforwardly to deal with more general large
deviations events, defined in terms of arbitrary linear combinations 
of $\{n_i\}$, that is, events of the form
$\{(n_0,n_1,\ldots,n_{M-1}):~\sum_{i=0}^{M-1}u_in_i \ge M\cdot U\}$, where
$\{u_i\}_{i=0}^{M-1}$ are arbitrary deterministic constants. For a meaningful
definition of the asymptotic regime, one
has to define the behavior of the infinite sequence $u_0,u_1,u_2,\ldots$, as
was done concerning the infinite sequence of parameters $p_0,p_1,p_2,\ldots$. For the sake of
simplicity, we will assume $u_i$ to be a function of $p_i$, i.e.,
$u_i=u(p_i)$, for a certain given function $u:[0,1]\to\reals$. 
We are then considering large deviations events of the form
$\sum_{i=0}^{M-1}n_iu(p_i) \ge MU$.
In the example discussed in Section 2, $u(p)=p/\alpha$, 
so that the sum becomes $\sum_{i=0}^{M-1}n_i / \mu_i \ge MU$.
In the example of the ideal Bose gas, where $p_i=e^{-\beta\epsilon_i}$, the
energy constraint $\sum_in_i\epsilon_i > MU$ corresponds to
$u(p)=-\frac{1}{\beta}\ln p$ \footnote{Albeit, in this case, 
the corresponding constraint does not give rise to condensation.}. Similarly as in the earlier derivation, the
saddle--point equation is now given by
$$U=\int_0^{p_m}\frac{tu(t)z^{u(t)}g(t)\mbox{d}t}{1-tz^{u(t)}},$$
whose solution $z$ should be sought in the 
interval $[1,Z)$, where $Z \equiv\inf_{p\in[0,p_m]}\{p^{-1/u(p)}\}$.
Thus, for
$$U > U_c\equiv
\int_0^{p_m}\frac{tu(t)Z^{u(t)}g(t)\mbox{d}t}{1-tZ^{u(t)}},$$
condensation takes place, provided that $U_c < \infty$. 

Several comments are now in order:
\begin{enumerate}
\item It is a simple exercise to show that in two dimensions
and above, an ordinary black--body would undergo a condensation 
when a constraint on the total number of photons is considered. This is evident by identifying $g(t)$ as the density of states of the photons, $U$ as the density of photons in the event considered and $p_i=e^{-\beta \varepsilon_i}$ with $\varepsilon_i$ the energy of a photon in mode $i$.

\item Different constraints can lead to condensates in different places. 
For example, assume that the hopping rates in the model of Section 2 
are ordered so that the slowest site is at site $i=0$ 
and the fastest is at site $i=M-1$. By looking at a constraint on $U$, one obtains 
a condensation at site $i=0$. However, if one looks at a constraint on the
quantity $Q=\sum_{i \leq M/2} (\mu_i - \mu_{M/2})^{\psi} n_i$, 
one can obtain a condensation at $i=M/2$ if $\psi$ is large enough.
	
\item In the ordinary BEC, where $u(t)\equiv 1$, the critical density could be finite only if $g(p_m)=0$
and $\lim_{t\uparrow p_m} g(t)/(p_m-t)^\chi$ is positive and finite
for some $\chi > 0$. In the more general case
considered now, there are choices for non--negative functions $u(t)$
such that $U_c < \infty$ even if $g$ does not vanish at $p_m$.
What counts is the rate
at which the denominator of the integrand, $1-tz_0^{u(t)}$,
tends to zero as $t\to p_m$. If $1-tz_0^{u(t)}$ behaves like
$|t-p_m|^\chi$ in the neighborhood of $p_m$, for some $0< \chi < 1$,
and $g(t)$ is continuous and finite at $t=p_m$,
then $U_c < \infty$. This in turn 
is possible because then the corresponding $u(t)$
would behave like $\log[(1-|t-p_m|^\chi)/t]$, which is positive in the
neighborhood of $p_m$.
\end{enumerate}

Having covered the single constraint problem, 
we now turn to the more interesting case where two constraints are considered simultaneously. 
Note that the analogy with a change of an ensemble is much weaker here. 
When considering large deviations, there is a freedom to choose any combination of constraints, so that in contrast 
to the usual statistical physics, the phase diagrams can have 
arbitrary dimensions.

\section{Two Constraints}

\begin{figure}[ht]
\hspace*{1cm}\input{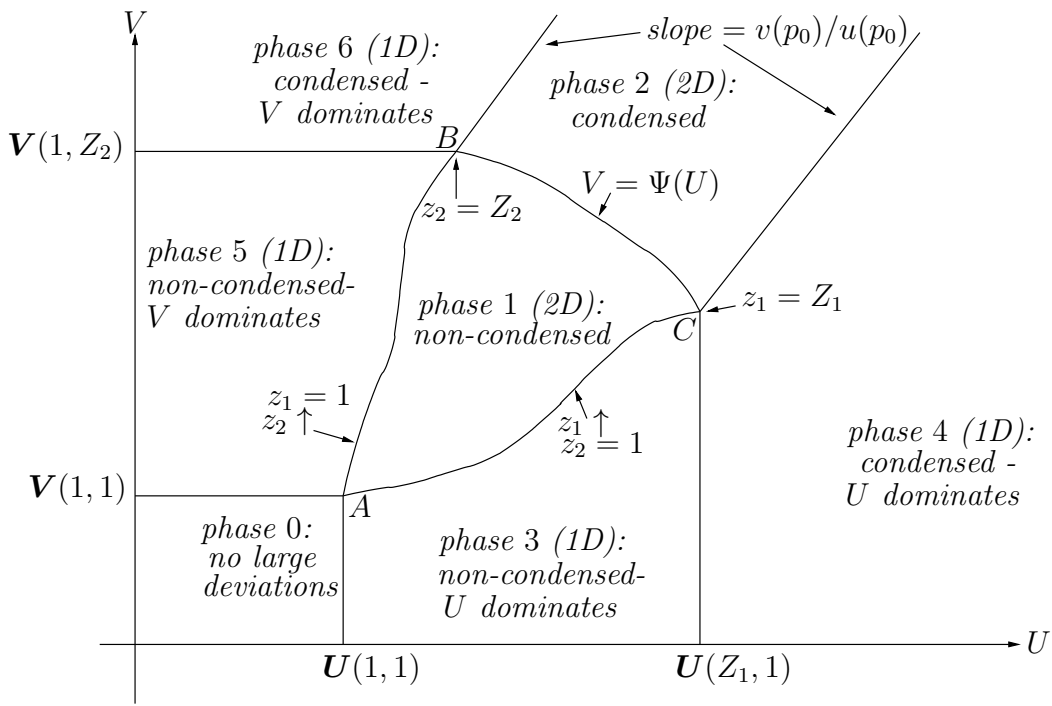}
\caption{\small Phase diagram in the $U-V$ plane. Note that each one the points $A$,
$B$ and $C$ is the meeting points
of four different phases.}
\label{uvplane}
\end{figure}

Having viewed the BEC from a large deviations perspective, it is instructive to
further extend the scope and consider the joint large deviations 
behavior of two events or more. Consider the rate function of two
joint events
$$\mbox{Pr}\left\{
\sum_{i=0}^{M-1}u_in_i \geq M U,~
\sum_{i=0}^{M-1}v_in_i \geq M V\right\},$$
where, once again, for the sake of 
simplicity, we assume that $u_i$ and $v_i$ depend on $i$ only via
$p_i$, i.e., $u_i=u(p_i)$ and $v_i=v(p_i)$ for certain given functions
$u(\cdot)$ and $v(\cdot)$. We confine ourselves to the case 
where the functions $u(\cdot)$ and $v(\cdot)$ are non--negative. 
This accommodates the examples discussed earlier in Section 2.
Denoting $\bX=\{
\sum_{i=0}^{M-1}u(p_i)n_i \geq M U,~
\sum_{i=0}^{M-1}v(p_i)n_i \geq M V\},$
and applying a two--dimensional Chernoff bound, we have:
\begin{eqnarray}
& &\mbox{Pr}\left\{ {\bf X} \right\}\nonumber\\
&\le&
\left< z_1^{\sum_{i=0}^{M-1}u(p_i)n_i -M U}\cdot
z_2^{\sum_{i=0}^{M-1}v(p_i)n_i -M V}\right>~~~~~~~z_1\ge
1,~z_2\ge 1\nonumber\\
&=&z_1^{-M U}z_2^{-M
V}\prod_i\left[(1-p_i)\sum_{n_i=0}^\infty[p_i z_1^{u(p_i)} z_2^{v(p_i})]^{n_i}\right]\nonumber\\
&=&z_1^{-M U} z_2^{-M
V}\prod_i\left[\frac{1-p_i}{1-p_i z_1^{u(p_i)} z_2^{v(p_i)}}\right]~~~~~~~\forall
i~ z_1^{u(p_i)} z_2^{v(p_i)}p_i < 1\nonumber\\
&=&\exp\left\{-M\left[U\ln
z_1+V\ln z_2-\frac{1}{M}\sum_{i=0}^{M-1}
\ln\left(\frac{1-p_i}{1-p_i\ z_1^{u(p_i)}
z_2^{v(p_i)}}\right)\right]\right\}.
\label{probtwo}
\end{eqnarray}
Again, the limitation $z_1 \geq 1$ and $z_2 \geq 1$ ensures 
that when we look at events where $U$ and $V$ take on values smaller than the
expectations $\lim_{M\to\infty}\frac{1}{M}\sum_iu(p_i)\left<n_i\right>$
and $\lim_{M\to\infty}\frac{1}{M}\sum_iv(p_i)\left<n_i\right>$, respectively,
the rate function would vanish. As before, to derive the rate function, 
we maximize the expression in the square brackets (which is a saddle--point
analysis)
by equating its partial derivatives with respect to $z_1$ and $z_2$ to zero. In the
thermodynamic limit, we get the following two equations with the two unknowns
$z_1$ and $z_2$:
\begin{eqnarray}
U&=&\bU(z_1,z_2)\equiv \int_0^{p_m}\frac{tu(t)z_1^{u(t)}z_2^{v(t)}g(t)\mbox{d}t}{1-tz_1^{u(t)}z_2^{v(t)}} \nonumber \\
V&=&\bV(z_1,z_2)\equiv\int_0^{p_m}\frac{tv(t)z_1^{u(t)}z_2^{v(t)}g(t)\mbox{d}t}{1-tz_1^{u(t)}z_2^{v(t)}}
\label{twofug}
\end{eqnarray}
where as before, $p_m$ is the maximum of $\{p_i\}$, which is again assumed to
be attained at $i=0$ for all $M$.
In analogy to usual BEC, $z_1$ and $z_2$ are jointly limited by the inequality $\sup_t[t z_1^{u(t)} z_2^{v(t)}] < 1$,
or equivalently,
\begin{equation}
\sup_{0\le t\le p_m}[u(t)\ln z_1+v(t)\ln z_2+\ln t]< 0.
\label{bound}
\end{equation}
In the sequel, we will refer to the following notation: For a given $z_1$, let
$\phi(z_1)$ be the supremum of the values of $z_2$ that do not violate eq.\
(\ref{bound}), and 
let $\calA=\{(z_1,z_2):~z_1\ge 1,~z_2\ge 1,~z_2 < \phi(z_1)\}$.
We now use the eqs.\ (\ref{twofug}) and (\ref{bound}) to derive the 
phase diagram for the large deviations rate function. For convenience, 
the final results are summarized towards the end of this section. The phase diagram, shown in Fig.\ \ref{uvplane}, has seven different phases. 

\vspace{0.2cm}

\noindent {\bf Phase 0: not a rare event}. The first, trivial, phase 
occurs when both $U$ and $V$ take on values below the expectations, $\bU(1,1)$ and
$\bV(1,1)$, respectively. This is the region where the events are not rare and so, $J(U,V)=0$. 

\vspace{0.2cm}

\noindent {\bf Phase 1: no condensation}. This phase is analogous to 
the non--condensed phase of the single event. Here, as long as the pair $(U,V)$
falls in a region for which
the equations 
$$U=\bU(z_1,z_2);~~~~
V=\bV(z_1,z_2)$$
have a solution $(z_1,z_2)\in\calA$, then one may substitute 
this solution into the Chernoff bound and obtain the rate function,
which in the thermodynamic limit is given by
$$J(U,V)=
U\ln z_1+V\ln z_2-
\int \mbox{d}tg(t)\ln\left[\frac{1-t}
{1-tz_1^{u(t)}z_2^{v(t)}}\right].$$
This phase is the image (under the transformation defined by the pair of
equations
$U=\bU(z_1,z_2)$, $V=\bV(z_1,z_2)$) of the set $\calA$ in the $z_1$--$z_2$
plane: It is surrounded by three
curves that connect the points $A$, $B$ and $C$ in Fig.\ \ref{uvplane}. 
The curve $A$--$B$ corresponds to the collection of points where $z_1=1$,
while $z_2$ varies from $1$
(point $A$) up to its maximum allowed value $z_2=\phi(1)\equiv Z_2$ (point $B$).
Similarly, the curve $A$--$C$ corresponds to $z_2=1$ and $z_1$ varying from
$1$ to $\phi^{-1}(1)\equiv Z_1$. Finally, the curve $B$--$C$ corresponds to the
curve $z_2=\phi(z_1)$, where as $z_1$ increases from $1$ to $Z_1$, $\phi(z_1)$
decreases from $Z_2$ to $1$. The image of the latter curve in the $U$--$V$
plane will be denoted by
$V=\Psi(U)$. Note that Fig.\ \ref{uvplane} assumes that the curve $A$--$B$
is above the curve $A$--$C$, namely, that
$\bU(1,z_2)\ge \bU(z_1,1)$ implies
$\bV(1,z_2)\ge \bV(z_1,1)$. In the appendix, we prove that
this is indeed always the case.

\vspace{0.2cm}

\noindent {\bf Phase 2: two--dimensional condensation}. We now consider the regime above
the curve $V=\Psi(U)$. Let us use the short--hand notation for the values that $U$ and $V$ take along the curve
$$\tilde{\bU}(z_1) \equiv \bU(z_1,\phi(z_1))$$
$$\tilde{\bV}(z_1) \equiv\bV(z_1,\phi(z_1)).$$
We assume, for the moment, that they are both 
finite for all $z_1\in[1,Z_1]$ and that $p_m$, the achiever of
$\sup_t[tz_1^{u(t)}[\phi(z_1)]^{v(t)}]$, is independent of $z_1$ 
(see the discussion in the sequel). 
Both these conditions are trivially met, for example, 
in the model and constraints discussed in Section 2. Let $(U,V)$ be 
a point above the curve $V=\Psi(U)$. The calculation of the rate function 
is somewhat more involved than the single constraint case. 
To describe it, we need to give values for both $z_1$ and $z_2$. To do this, we note that
in analogy to usual BEC, we have:
$$U-\tilde{\bU}(z_1)=\lim_{M\to\infty}\left[\frac{1}{M}
\cdot\frac{p_mu(p_m)
z_1^{u(p_m)}z_2^{v(p_m)}}{1-p_m z_1^{u(p_m)} z_2^{v(p_m)}}\right]$$
and similarly,
$$V-\tilde{\bV}(z_1)=\lim_{M\to\infty}\left[\frac{1}{M}
\cdot\frac{p_mv(p_m)
z_1^{u(p_m)}z_2^{v(p_m)}}{1-p_m z_1^{u(p_m)} z_2^{v(p_m)}}\right] \;.$$
As in the ordinary BEC, where a prescription has to be specified for 
how the fugacity approaches the condensation value in the condensed phase, 
these equations essentially give a prescription for taking the values of the fugacities 
to a point where $\sup_t[tz_1^{u(t)}[\phi(z_1)]^{v(t)}]=1$, 
as the thermodynamics limit is taken. Using these, we see that
$$\frac{V-\tilde{\bV}(z_1)}
{U-\tilde{\bU}(z_1)}=\frac{v(p_m)}{u(p_m)}\;.$$
This equation specifies, given a point $(U,V)$ above the curve $V=\Psi(U)$, the choice of $z_1$, which we shall
denote by $z_1^*$, and hence also the choice of $z_2$,
which is $z_2^*=\phi(z_1^*)$. The large deviations event is
dominated by the state corresponding to $t=p_m$.
Thus, the rate function is given by
$$J(U,V)=
U\ln z_1^*+V\ln z_2^*-\int g(t)\mbox{d}t\ln
\left[\frac{1-t}{1-t(z_1^*)^{u(t)}(z_2^*)^{v(t)}}\right].$$
It must be kept in mind, however, that this solution is not applicable
to all points $(U,V)$ above the curve $V=\Psi(U)$. To understand the
limitation, it is instructive to look at the geometric interpretation
of the above equation for $z_1^*$: The expression
$[V-\tilde{\bV}(z_1)]/
[U-\tilde{\bU}(z_1)]$ is the slope of the straight line connecting
the point $(U,V)$ to the point
$(\tilde{\bU}(z_1),\tilde{\bV}(z_1))$ on
the curve $V=\Psi(U)$, and the equation
tells us that this slope must be equal to $v(p_m)/u(p_m)$, which is
a given constant. Therefore, this solution is applicable only to
points $(U,V)$ above the curve $V=\Psi(U)$ which have the following property:
the straight line of slope $v(p_m)/u(p_m)$ that passes through $(U,V)$
must intersect the curve $V=\Psi(U)$ between points $B$ and $C$. 
The set of points with this property,
which corresponds to the region of two--dimensional condensation
is limited by the curve $V=\Psi(U)$
(between $B$ and $C$) and the two parallel straight lines of slope
$v(p_m)/u(p_m)$,
passing through $B$ and $C$ (see Fig.\ \ref{uvplane}). 

\vspace{0.2cm}

\noindent {\bf Phase 3: non--condensed and dominated by the $U$--constraint}. 
The region below the curve $A$--$C$
(see Fig.\ \ref{uvplane}) is characterized by $z_2=1$ and $z_1 \geq 1$. 
The value of $z_2$ is fixed at unity since we are 
considering values of $V$ which are below the corresponding average value 
conditioned on the given value of $U$. 
This means that there is a non--condensate large deviations behavior that is dominated by
that of the constraint $\sum_i u(p_i)n_i \ge M U$ alone.
In other words, the other event, 
$\sum_i v(p_i)n_i \ge M V$, has no impact. The 
rate function is given by minimizing the term in the square brackets in eq.\
(\ref{probtwo}) 
with $z_2=1$. Denoting the obtained value of $z_1$ by $z_1^*$, the rate function is
given by
$$J(U,V)=U\ln z^*_1-\int_0^{p_m} \mbox{d}tg(t) \ln\left[\frac{1-t}
{1-t (z_1^*)^{u(t)}}\right].$$
This phase is bounded on the right by a vertical line (see Fig.\
\ref{uvplane}), where the constraint $\sum_i u(p_i)n_i \ge M U$ condenses with $z_2=1$.

\vspace{0.2cm}

\noindent {\bf Phase 4: condensed and dominated by the $U$--constraint}. 
Following the reasoning of phase 3, the region below the straight line 
of slope $v(p_m)/u(p_m)$, passing via $C$, is the corresponding condensed 
phase of this single event $\sum_i u(p_i)n_i \ge M U$ (one-dimensional
condensation), where the constraint $\sum_i v(p_i)n_i \ge M V$ has no impact. 
The upper bound on the phase can be inferred by noting that on the line
emerging from point $C$ in the figure, $z_2=1$. 

The last two phases can be inferred from a symmetry consideration, where the
two constraints interchange their roles.

\vspace{0.2cm}

\noindent {\bf Phase 5: non--condensed and dominated by the $V$--constraint}. 
See the discussion for phase 3.

\vspace{0.2cm}

\noindent {\bf Phase 6: condensed and dominated by the $V$--constraint}. See the discussion for phase 4.


Let us examine now more closely the assumption that $\tilde{\bU}(z_1)$ and
$\tilde{\bV}(z_1)$ 
are both finite for a continuum of values of $z_1$. 
In the two--dimensional case considered now, this issue is more involved
than in the one--dimensional case: In the one dimensional case, the
relevant integral, computed at the maximum allowed value of the
fugacity parameter $z$, may be finite if the density
$g(t)$ vanishes at $t=p_m$ (the achiever of $\min_p
p^{-1/u(p)}$), and tends to zero sufficiently rapidly as $t\to p_m$.
By contrast, in the two--dimensional case considered now,
the achiever
of $\sup_t\{t z_1^{u(t)}[\phi(z_1)]^{v(t)}\}$, 
may depend, in general, on $z_1$,
and it is inconceivable to expect $g(t)$ to vanish at all these values
of $t$, which may form a continuum.
(In fact, if $g(t)=0$ for an
interval, then this interval has no contribution to the integrals altogether.)
Nonetheless, there is a class of special cases
where this situation does not arise -- the cases where the maximizing
value of $t$ turns out to
be independent of $z_1$: For example,
if $u(t)$ and $v(t)$ are both monotonically
non-decreasing, then $\sup_t\{tz_1^{u(t)}[\phi(z_1)]^{v(t)}\}$
is always achieved at $t=p_m$, independent of $z_1$, 
where now $p_m$ is again the maximum value of $p$ across the support of the
density $g(t)$. In this
case, as in the one--dimensional case,
if $g(t)\to 0$ as $t\uparrow p_m$ sufficiently rapidly,
then $\tilde{\bU}(z_1)$ and $\tilde{\bV}(z_1)$ are both
finite, and then for large enough $U$ and $V$, there is a condensation
at the state corresponding to $p_m$, as explained above.
It should be noted, however, that the non--decreasing monotonicity of $u(t)$
and $v(t)$ is only a sufficient condition for $p_m$ to be independent of
$z_1$, not a necessary condition. For example, ignoring our previous 
assumption on the positivity of $u(t)$ and $v(t)$, if $u(t)\equiv 1$ and $v(t)=-\ln t$, this is still
true, although $v(t)=-\ln t$ is a decreasing function. 

To summarize, we have identified seven 
phases in the $U$--$V$ plane.
Denoting
$$\calJ(z_1,z_2,U,V)=U\ln z_1+V\ln z_2-
\int g(t)\mbox{d}t \ln\left[\frac{1-t}
{1-t z_1^{u(t)} z_2^{v(t)}}\right],$$
the rate function takes the following behaviors:
$$J(U,V)=\left\{\begin{array}{ll}
0 & \mbox{phase $0$}\\
\max_{z_1,z_2}\calJ(z_1,z_2,U,V) & \mbox{phase $1$}\\
\calJ(z_1^*,\phi(z_1^*),U,V) & \mbox{phase $2$}\\
\max_{z_1}\calJ(z_1,1,U,V) & \mbox{phase $3$}\\
\calJ(Z_1,1,U,V) & \mbox{phase $4$}\\
\max_{z_2}\calJ(1,z_2,U,V) & \mbox{phase $5$}\\
\calJ(1,Z_2,U,V) & \mbox{phase $6$}\end{array}\right.$$
It is interesting to compare this phase diagram with the one which would 
be obtained by considering the equality event
$$\mbox{Pr}\left\{ {\bf X} \right\} =\mbox{Pr}\left\{
\sum_{i=0}^{M-1}u(p_i)n_i = M U,~
\sum_{i=0}^{M-1}v(p_i)n_i = M V\right\}.$$
In this case, the values of both $z_1$ and $z_2$ 
would not be restricted to be larger than $1$. 
Therefore, all phase transitions associated with either 
$z_1=1$ or $z_2=1$ would disappear in this case. It is straightforward to see 
(similarly to the derivation of phase 1), that here we have two phases only: a
condensed phase and a non--condensed phase. In the $z_1$--$z_2$ plane, the set
$\calA$ is no longer limited by the inequalities $z_1\ge 1$ and $z_2\ge 1$,
but only the curve $z_2=\phi(z_1)$, whose image in the $U$--$V$ plane is now
the {\it entire} curve $V=\Psi(U)$, which is no longer limited by the points
$B$ and $C$. The region below this curve is the non--condensed phase and the
region above the curve is condensed. The condensation is always
two--dimensional in character.

Finally, it would be interesting to demonstrate that in certain situations,
the condensating state may jump abruptly as we move continuously in the
$U$--$V$ plane.
In the above discussion we made specific assumptions on the functions $g(t)$,
$u(t)$, and $u(t)$. In principle, it is possible to extend the calculation to 
cases where the achiever of
$\sup_t\{tz_1^{u(t)}[\phi(z_1)]^{v(t)}\}$ takes on any finite
number of values as $z_1$ varies between $1$ and $Z_1$,
and and that the density $g$ vanishes (and sufficiently rapidly) at
all these values of $t$. An interesting scenario arises, for example, in a variation of the above example,
defined by the choices $u(t)\equiv 1$ and $v(t)=-\alpha-\ln t$, where
$0< \alpha < -\ln p_m$, and where as before, $p_m$ is the maximum of $t$ across the
support of $g(t)$. In this case, it is easy to see that the achiever of 
$\sup_t\{tz_1^{u(t)}[\phi(z_1)]^{v(t)}\}$ is given by $p_m$ for $z_2 <
e$ ($z_1 > e^\alpha$), and by $p_\infty$, which is the minimum of $t$ across the support of $g(t)$, for
$z_2 > e$ ($z_1 < e^\alpha$). 
In other words, the condensing state jumps from $p_m$ to the
other extreme, $p_\infty$, at
the point $z_1=e^\alpha$ along the curve
$V=\Psi(U)$. 
In this case, the two--dimensional condensed phase
splits into three sub--phases. If we denote by $D$ the point
corresponding to $z_1=e^\alpha$ along the curve $V=\Psi(U)$, then above
this curve, we see three different types of two--dimensional condensation (see
Fig.\ \ref{splitcond}):

\begin{figure}[ht]
\hspace*{3cm}\input{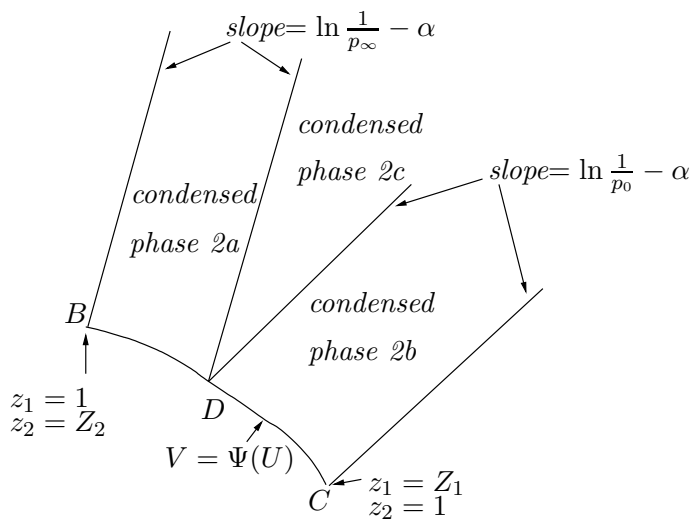}
\caption{\small Zoom--in on the two-dimensional condensed phase
in the example of $u(t)\equiv 1$ and $v(t)=-\alpha-\ln t$.}
\label{splitcond}
\end{figure}

\begin{enumerate}
\item
The region limited by the curve $B$--$D$ and two parallel straight lines
with slope $\ln(1/p_\infty)-\alpha$, passing through points $B$ and $D$ (phase 2a).
\item
The region limited by the curve $D$--$C$ and two parallel straight lines
with slope $\ln(1/p_0)-\alpha$, passing through points $D$ and $C$ (phase 2b). 
\item The
region in between 1 and 2 (phase 2c). The rate function for all points in
phase 2c is the same as in the point $D$.
\end{enumerate}

\section{Applications}

Our large deviations analysis focuses on events associated with linear
combinations pertaining to sequences of independent (but not
necessarily identically distributed) geometric random variables. Beyond
the obvious relevance of this model to the grand--canonical ensemble of
the ideal boson gas, as was mentioned earlier, there are quite a few
additional applications, which cover, not only the realm of statistical physics, but also that
of information
engineering models. We mentioned briefly some of these applications in the Introduction.
In this section, we discuss them in somewhat more detail.

\begin{figure}[ht]
\hspace*{3cm}\input{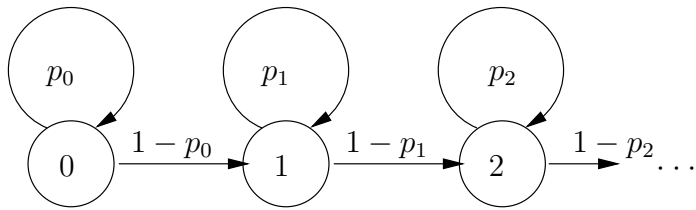}
\caption{\small State transition diagram for a one--way Markov chain.}
\label{owmc}
\end{figure}

The first application example is that of a {\it one--way Markov chain}
(a.k.a.\ {\it left--to--right} Markov chain, in the literature of speech
signal processing). A one--way Markov chain
is defined by an ordered set of states ($0,1,2,\ldots$),
where the only allowed transitions from each state $i$ are
the self--transition ($i\to i$) -- with probability
$p_i$, and a transition to the next state ($i\to i+1$) -- with probability
$1-p_i$, $i=0,1,2,\ldots$ (see Fig.\ \ref{owmc}). 
Clearly, every sequence generated by a one--way Markov
chain, as defined, is composed of $n_0$ self--transitions of state $0$,
followed
by $n_1$ self--transitions of state $1$, followed in turn by $n_2$
self--transitions of state $2$, and so on, where $n_0,n_1,n_2,\ldots$ are
independent, geometric random variables with parameters $\{p_i\}$.

Therefore, it is clear that this model falls within our framework.
The one--way Markov chain is a very useful model in a variety of application
areas of information
system models.
A few examples are
hidden Markov modeling of speech signals (see, e.g., \cite{Rabiner89}
and references therein), the segmentation of signals, such as those
that govern the evolution of the fading process of a communication
channel (or channels that ``heat up'' \cite{KLS08}), the segmentation
of electrocardiographic signals (see, e.g., \cite{MCN04}), beat tracking in
audio signals (see, e.g., \cite{SK08}), and even handwritten text recognition
\cite{MCN04}. 

The interest in the large deviations behavior of linear combinations
of $\{n_i\}$ is not difficult to justify, in the context of one--way
Markov chains.
Consider the problem of lossless data compression of the sequence of random
variables $n_0,n_1,\ldots$. An elementary result in Information Theory
(see, e.g., \cite{CT06}) tells that 
the optimum code length (in bits) of the compressed version of each $n_i$ is
given by
$$\ell_i(n_i)=-\log
P(n_i)=-\log[(1-p_i)p_i^{n_i}]=n_i\log(1/p_i)+\log[1/(1-p_i)],$$
which is an affine function of $n_i$. The large deviations event
$\sum_i\ell_i(n_i)\ge N=MU$ is the event that the total code length
would exceed the limit of $N$. If $N$ designates the size of a buffer
in which the compressed data is stored (in order to monitor the bit rate),
then this event has the meaning of a buffer overflow, whose consequence is that
information is lost. We would like, of course, to keep the probability of
such an event as small as possible.

Another application where independent geometric random variables naturally
arise, is in queuing theory.
An M/M/1 queue (see, e.g., \cite{PRS76}) is a common model of a queue
according to
which the arrivals of customers is a Poisson process of rate $\lambda$,
the service is based on the principle of first come -- first served (FCFS),
and the service time for each customer is distributed exponentially with rate
$\mu$.
As long as $\lambda < \mu$, the queue is stable (does not diverge) and the
steady--state
distribution of the number of customers in the queue is geometric
with parameter $p=\lambda/\mu$, which is called the {\it utilization} of
the queue.
Jackson's theorem \cite{Jackson57}
extends this to an open network (a.k.a.\ a Jackson network) of $M$ queues,
which means that: (i) any external arrival to any given node is a Poisson
process, (ii) a customer completing service at queue $i$ either joins another
queue $j$ with probability $p_{ij}$ or leaves the system with probability
$1-\sum_j p_{ij}$, which is non--zero for at least one queue, and (iii) all
utilization parameters $p_i$
are less than 1. Jackson's theorem tells that the
steady--state joint probability distribution of
the queue lengths is given by a product of individual
geometric distributions with parameters $\{p_i\}$. 
A special case of a queuing network was considered in Section 2.

In the context of queuing
networks, BEC means that one of the queues, the one with the highest
utilization, becomes responsible for a bottleneck (or a traffic jam)
--  a linear fraction of the total number of
customers spend their time in that queue due to the inefficient
performance of the server of this queue relative to the arrival rate.
When applied to queuing networks,
our large deviations results mean that we
identified BEC in an open (Jackson) network
and in addition, we have characterized the rate function, as well as
the phase transitions associated with it. 
Moreover, since we are allowing large
deviations
events pertaining to arbitrary linear combinations of $\{n_i\}$, one
natural application example, as already discussed, is the large deviations behavior of
$\sum_i n_i/\mu_i$ (with $\mu_i$ being the
rate through queue no.\ $i$), which is
a reasonable estimate of the total waiting time for a customer who visits all
queues.

There are, of course, other network models that are known to admit a
product--form
steady--state distribution. One of them is
the closed--network version of the Jackson network, called
the {\it Gordon--Newell network}
\cite{Robert03},\cite{Walrand83}. The only difference
between the Gordon--Newell network and the
Jackson network is that the former is a closed network
(unlike a Jackson network which is open), i.e., there is no
external supply of customers and no departures from the system, and so, the
total number of customers is fixed. The steady--state distribution for
the Gordon--Newell network is exactly analogous to the canonical
Bose--Einstein
distribution, and hence it exhibits BEC under certain conditions, as was
observed already in earlier work, cf.\ e.g.,
\cite{MY93} and \cite{FL96}.

The Gordon--Newell theorem appears to be a special case of
results concerning product forms of steady--state distributions
in classes of models, such as the zero--range process (ZRP)
(see, e.g., \cite{Evans96}, \cite{Evans00} and references therein),
that are studied in the statistical physics literature. According to the ZRP
model, particles (customers) that lie in an array of sites (a lattice,
or more generally, the nodes of a
certain graph), may hop from
one site (queue) to another, and may pile up, according to certain rules
(see, e.g., the example discussed in Section 2).
Jackson's theorem, however, does not seem to be
directly derivable as a special case since it pertains to an open network.
A subsequent paper by Jackson \cite{Jackson63}
allows state--dependent service times and it seems to include the ZRP
model as a special case.

\section*{Acknowledgment} Useful discussions with Martin Evans are gratefully
acknowledged.

\section*{Appendix}

In this appendix, we prove that $\bU(1,z_2)\ge \bU(z_1,1)$ implies
$\bV(1,z_2)\ge \bV(z_1,1)$, which means that the $A$--$B$ curve
in Fig.\ \ref{uvplane} lies above the $A$--$C$ curve.

Consider the function
$$f(x)=\frac{tg(t)e^x}{1-te^x},$$
where $t$ is a parameter taking values in the range
where the denominator is strictly positive. For a given $t\ge 0$,
this function is clearly monotonically non--decreasing in $x$.
Therefore, for all $t$:
$$[u(t)\ln z_1-v(t)\ln z_2]\cdot
[f(u(t)\ln z_1)-f(v(t)\ln z_2)]\ge 0.$$
Integrating over $t$, we get:
\begin{eqnarray}
0&\le&\int\mbox{d}t
[u(t)\ln z_1-v(t)\ln z_2]\cdot
[f(u(t)\ln z_1)-f(v(t)\ln z_2)]\nonumber\\
&=&\left[\int\frac{tu(t)z_1^{u(t)}g(t)\mbox{d}t}{1-t z_1^{u(t)}}-
\int\frac{tu(t)z_2^{v(t)}g(t)\mbox{d}t}
{1-t z_2^{v(t)}}\right]\cdot\ln z_1+\nonumber\\
& &\left[\int\frac{tv(t)z_2^{v(t)}g(t)\mbox{d}t}{1-tz_2^{v(t)}}-
\int\frac{tv(t)z_1^{u(t)}g(t)\mbox{d}t}{1-tz_1^{u(t)}}\right]\cdot\ln z_2\nonumber\\
&=&[\bU(z_1,1)-\bU(1,z_2)]\cdot\ln z_1+
[\bV(1,z_2)-\bV(z_1,1)]\cdot\ln z_2.
\end{eqnarray}
Since the first bracketed term of the last expression
is non--positive (by hypothesis)
and since $\ln z_1\ge 0$ and $\ln z_2\ge 0$,
the second bracketed term must be
non--negative, which proves the argument.

\end{document}